\documentclass[11pt,a4paper]{article}
\usepackage{color}
\usepackage{epsfig}
\usepackage[affil-it]{authblk}
\usepackage[colorlinks = true,
            linkcolor = blue,
            urlcolor  = blue,
            citecolor = blue,
            anchorcolor = blue]{hyperref}
\usepackage{graphicx}
\usepackage{dcolumn}
\usepackage{amsmath, amssymb}
\usepackage{textcomp}
\usepackage{float}
\usepackage{subfig}
\usepackage{fancyhdr}
\floatstyle{boxed}
\usepackage[]{geometry}
\title{\bf{The Raychaudhuri equation for a quantized timelike geodesic congruence} }
\author[,1]{Shibendu Gupta Choudhury\thanks{Email: sgc14ip003@iiserkol.ac.in}}
\author[,1]{Ananda Dasgupta \thanks{Email: adg@iiserkol.ac.in}}
\author[,1]{Narayan Banerjee \thanks{Email: narayan@iiserkol.ac.in}}
\affil[1]{Department of Physical Sciences, Indian Institute of Science Education and Research Kolkata, Mohanpur, Nadia 741246, India}
\date{}

\begin{document}
\sloppy
\maketitle
\begin{abstract}
A recent attempt to arrive at a quantum version of Raychaudhuri's equation is looked at critically. It is shown that the method, and even the idea, has some inherent problems. The issues are pointed out here. We have also shown that it is possible to salvage the method in some limited domain of applicability. Although no generality can be claimed, a quantum version of the equation should be useful in the context of ascertaining the existence of a  singularity in the quantum regime. The equation presented in the present work holds for  arbitrary $n+1$ dimensions. An important feature of the Hamiltonian in the operator form is that it admits a self-adjoint extension quite generally. Thus, the conservation of probability is ensured.
\end{abstract}

\maketitle

\section{Introduction}\label{sec1}
Although General Relativity is the most successful theory of gravity so far, the existence of a singularity in a classical spacetime is inevitable in this theory as 
demonstrated by the Penrose-Hawking singularity theorems\cite{penrose, hawking}. Physical laws break down and the spacetime geometry is pathological at a singularity. 
Therefore, for a complete physical description of the spacetime structure, these singularities must be looked at more critically. 

A general expectation is that quantum effects which come into the picture in strong gravity regime may alleviate this problem.  As there is no universally accepted quantum 
theory of gravity, quantum effects have to be explored in specific gravitational systems. Prescriptions for resolving singularities are there in the literature, such as in string 
theoretical framework\cite{niz, leh1, leh2, green, cope}, loop quantum gravity\cite{casa,lang,batt,asht0,agu} and other different approaches to quantum 
gravity\cite{awad,trug,gara,oriti,hof}. But all of them suffer from some issue or the  other\cite{fermilab,rosz,nicol}.

A key ingredient in the classical singularity theorems is the focusing theorem which is a consequence of the Raychaudhuri equation\cite{rc,ehlers}. Application of Raychaudhuri's 
equation in quantum settings may be useful in deciding the existence or resolution of the singularities at the quantum level.  A quantum version of Raychaudhuri's equation was 
developed by Das\cite{saurya} using Bohmian trajectories and it was argued that focusing of geodesics does not occur when the quantum potential is included. The implications of 
this equation in  cosmological and black hole  spacetimes were discussed in \cite{ali1, ali2}. The Raychaudhuri equation has been useful in showing the avoidance of 
singularities in loop quantum cosmology\cite{bojo, asht, singh, li}. This is possible due to the existence of repulsive terms which arise due 
to quantum effects. Similar analysis in other contexts can be found in the references \cite{ashtbojo, burg, moti}. The Raychaudhuri equation has also been studied in the spacetime 
described by the qmetric\cite{sumanta}. This qmetric includes a zero point length by construction\cite{kothaw, paddykotha}. In \cite{sumanta} it has been established that this 
existence of the zero point length can help us avoid geodesic convergence. 

The idea of presenting a geodesic congruence as a dynamical system was suggested by Alsaleh, Alasfar, Faizal and Ali\cite{alsa}. Their method is an attempt to reveal what 
a \textit{back to basics} endeavour can do. They identify a suitable dynamical variable $\rho$, write down Raychaudhuri's equation in terms of this variable, construct a 
Lagrangian so that the Raychaudhuri equation is arrived at as the Euler-Lagrange equation from that Lagrangian. Now the quantization scheme is straightforward. One has to write 
down the momentum ($\Pi$) canonically conjugate to the proposed dynamical variable, construct the classical Hamiltonian, promote $\rho$ and its conjugate momentum to operators 
such that the canonical quantization scheme ($[\hat{\rho},\hat{\Pi}]=i\hbar$) is satisfied and thus write down the Hamiltonian \textit{operator}. The Schr\"odinger equation 
follows. Unfortunately one equation in their work contains an error, and the resulting quantization is not quite correct in general. Their scheme works only in $(2+1)$ dimensions. 
Also there are other deep-rooted problems in the scheme, which are pointed out in this paper.

The motivation of the present work is to find a ``correct'' quantum version of the Raychaudhuri equation following the approach as that of 
Alsaleh \textit{et al}\cite{alsa}. We  arrive at an equation which is valid for an arbitrary dimension. This is done with the help of a more rigorous approach by using the 
well-known Helmholtz conditions\cite{davis1,davis2,douglas,casetta,cramp,nigam} regarding the Lagrangian formulation of a problem. This helps us in writing down a viable 
Lagrangian as the starting point. We also point out other more serious problems, which may not be cured so easily. 

The perspective is to find a quantum version of geodesic flow equation primarily following basic canonical approach very similar to the formulation of Wheeler-DeWitt quantization 
scheme. Certainly the idea was to obtain a full quantization of the geodesic congruences. As we will see later, the scheme practically works for some simple minisuperspaces. This is appropriate as a starting point, for the reason that in a gravitational system, the quantum regime is normally investigated in a minisuperspace. A quantum description is unavoidable 
in the high energy scale, i.e., beyond the Planck scale of energy. {Observational signature will have to be ascertained by the analysis of primordial perturbations, and 
does not feature in the present work.}

The paper is organized as follows. In section \ref{sec2} we briefly discuss the Helmholtz conditions regarding the inverse problem of the Lagrangian formulation of a dynamical 
system. In section \ref{sec3} the idea of considering geodesic congruence as a dynamical system is discussed and it is shown that one can generalize this concept to arbitrary 
dimensions. This section also contains a caveat regarding the scheme and discusses about the limited area of possible application of the scheme. Section \ref{sec4} describes 
the canonical quantization of the system. The issues of operator ordering and self-adjoint extension(s) of the quantum Hamiltonian are also discussed in this section. The final 
section \ref{sec5} contains some discussions and concluding remarks. {Some possible  arena for applications are also discussed in the final section.}

\section{The Helmholtz conditions}\label{sec2}
In this section, we briefly review the Helmholtz conditions\cite{davis1,davis2,douglas,casetta,cramp,nigam} regarding the Lagrangian formulation of a system. Let us consider a 
system with $d$ degrees of freedom. This system is described by $d$ second order differential equations of the form,
\begin{equation}\label{eom}
 F_i(t,x_j,\dot{x}_j,\ddot{x}_j)=0,
\end{equation}
where dot denotes derivative with respect to $t$ and $i,j=1,2,...,d$.

The necessary and sufficient conditions which must be satisfied by equation \eqref{eom} for being the
Euler-Lagrange equation corresponding to a Lagrangian $L(t,x_j,\dot{x}_j)$, are known as the Helmholtz conditions. These conditions are \cite{nigam}:
 \begin{eqnarray}\label{h1}
  \frac{\partial F_i}{\partial \ddot{x}_j}&=&\frac{\partial F_j}{\partial \ddot{x}_i},\\ \label{h2}
  \frac{\partial F_i}{\partial {x_j}}-\frac{\partial F_j}{\partial {x_i}}&=&\frac{1}{2}\frac{\mathrm{d}}{\mathrm{d}t}\left(\frac{\partial F_i}{\partial \dot{x}_j}-\frac{\partial F_j}{\partial \dot{x}_i}\right),\\ \label{h3}
  \frac{\partial F_i}{\partial \dot{x}_j}+\frac{\partial F_j}{\partial \dot{x}_i}&=& 2\frac{\mathrm{d}}{\mathrm{d}t}\left(\frac{\partial F_j}{\partial \ddot{x}_i}\right),
 \end{eqnarray}
 for all $i,j=1,2,...,d$.
 
 In the next section we will see how one can utilize these conditions in the context of the representation of a geodesic congruence as a dynamical system.
 
\section{Geodesic congruence as a dynamical system}\label{sec3}

We consider a hypersurface orthogonal timelike geodesic congruence in an $(n+1)$-dimensional spacetime. Let $h_{\alpha\beta}$ be the induced metric on the $n$-dimensional 
hypersurface which is orthogonal to the timelike unit velocity vector $u^\mu$ of the congruence. If the congruence is considered as a dynamical system, one can define the 
dynamical degree of freedom as in  \cite{alsa},
\begin{equation}
 \rho(\lambda)=\sqrt{h},
\end{equation}
where $h=\det(h_{\alpha\beta})$ and $\lambda$ is the affine parameter.

The dynamical evolution of $h$ is given by\cite{poisson},
\begin{equation}
 \frac{1}{\sqrt{h}}\frac{\mathrm{d}\sqrt{h}}{\mathrm{d}\lambda}=\theta,
\end{equation}
where $\theta=\nabla_\mu u^\mu$ is the expansion scalar of the congruence.
Therefore, we have,
\begin{equation}\label{rho}
 \rho ^\prime\equiv\frac{\mathrm{d}\rho}{\mathrm{d}\lambda}=\rho\theta,
\end{equation}
where a prime indicates a differentiation with respect to the affine parameter $\lambda$. It should be pointed out here that there is an erroneous extra factor 
of $\frac{2}{n}$ on the right hand side of this equation in the work of Alsaleh \textit{et al}\cite{alsa}. This reduces to the correct result only for $n=2$.

In this case, the Raychaudhuri equation which dictates the evolution of the congruence,  is given by\cite{rc, ehlers},
\begin{equation}\label{raych}
 \frac{\mathrm{d}\theta}{\mathrm{d}\lambda}+\frac{1}{n}\theta^2+2\sigma^2+\mathcal{R}=0,
\end{equation}
where $2\sigma^2=\sigma_{\alpha\beta}\sigma^{\alpha\beta}$ and $\sigma_{\alpha\beta}=\nabla_{(\nu}u_{\mu)}-\frac{1}{n}h_{\alpha\beta}\theta$ is the shear tensor. 
$\mathcal{R}=R_{\mu\nu}u^\mu u^\nu$ with $R_{\mu\nu}$ being the Ricci tensor. As the congruence is hypersurface orthogonal, 
the rotation tensor $\omega_{\mu\nu}=\nabla_{[\nu}u_{\mu]}=0$\cite{poisson}.

Using equation \eqref{rho}, Raychaudhuri's equation \eqref{raych} can be written as,
 \begin{equation}\label{raycl}
  \frac{\rho^{\prime\prime}}{\rho}+\frac{{\rho^\prime}^2}{\rho^2}\left(\frac{1}{n}-1\right)+2\sigma^2+\mathcal{R}=0.
 \end{equation}
We want this equation as the Euler-Lagrange equation corresponding to a Lagrangian. For that, we compare this equation \eqref{raycl} with equation 
\eqref{eom} and we have,
\begin{equation}\label{fray}
 F=\frac{\rho^{\prime\prime}}{\rho}+\frac{{\rho^\prime}^2}{\rho^2}\left(\frac{1}{n}-1\right)+2\sigma^2+\mathcal{R}.
\end{equation}
It is easy to verify that with this $F$, conditions \eqref{h1} and \eqref{h2} are trivially satisfied but to satisfy the condition \eqref{h3} we need $n=2$.
So a Lagrangian formulation with this $F$, given by equation \eqref{fray}, is possible only in $(2+1)$-dimensions. This 
expression for $F$ is used in the work Alsaleh \textit{et al}\cite{alsa}. Therefore, their calculations are valid only for $n=2$.\\

Now, we discuss how one can generalize this idea to $(n+1)$-dimensions for an arbitrary $n$. To achieve this, we multiply  equation \eqref{raycl} by an integrating factor 
which in turn leads to the existence of a Lagrangian for the $(n+1)$-dimensional system under consideration. For a comprehensive discussion on this method we refer to the 
work of Casetta and Pesce\cite{casetta} and references therein. If we multiply $F$ in equation \eqref{fray} by $\rho^{\frac{2}{n}-1}$, then the new $\tilde{F}$,
\begin{equation}\label{ftil}
 \tilde{F}=\rho^{\left(\frac{2}{n}-1\right)}\left[\frac{{\rho ^{\prime\prime}}}{\rho}+\frac{{\rho ^\prime}^2}{\rho^2}\left(\frac{1}{n}-1\right)+2\sigma^2+\mathcal{R}\right],
\end{equation}
satisfies all the Helmholtz conditions if we demand that $2\sigma^2+\mathcal{R}$ is a function of $\rho$ only. Now, one can construct a Lagrangian as,
\begin{equation}\label{lagr}
 L=\frac{1}{2}\rho^{\left(\frac{2}{n}-2\right)}{\rho ^\prime}^2-V[\rho].
\end{equation}
 A variation of this Lagrangian, with respect to the dynamical variable $\rho$, yields
\begin{equation}\label{varia}
\begin{split}
 \delta \mathcal{L}= -\rho^{\left(\frac{2}{n}-1\right)}\left[\frac{{\rho ^{\prime\prime}}}{\rho}+\frac{{\rho ^\prime}^2}{\rho^2}\left(\frac{1}{n}-1\right)\right] \delta \rho-\delta V+ \frac{\mathrm{d}}{\mathrm{d}\lambda}\left(\rho^{\left(\frac{2}{n}-2\right)}\rho' \delta \rho\right).
 \end{split}
\end{equation}
 Therefore, to get the equation $\tilde{F}=0$ from the least action principle we need,
\begin{equation} \label{delta-v1}
 \frac{\delta V[\rho]}{\delta \rho}=\rho^{\left(\frac{2}{n}-1\right)}\left(2\sigma^2+\mathcal{R}\right).
\end{equation}
{It should be mentioned at this stage that $V[\rho]$ here is the potential corresponding to the dynamical system representing the congruence and has to be constructed using the gravitational field equations. It is not quite the potential, if any, 
 in the matter sector alone}.
 
 The Euler-Lagrange equation corresponding to this Lagrangian is,

\begin{equation}\label{el-eq}
 \frac{\partial L}{\partial \rho}=\frac{\mathrm{d}}{\mathrm{d}\lambda}\left(\frac{\partial L}{\partial \rho^\prime}\right).
\end{equation}

  If we now assume that $V$ is a function of $\rho$ alone (instead of being a functional) in equation \eqref{lagr}, equation \eqref{el-eq} implies,

\begin{equation}
 \rho^{\left(\frac{2}{n}-1\right)}\left[{\theta ^\prime}+\frac{1}{n}\theta^2+\mathcal{R}+2\sigma^2\right]=0.
\end{equation}
This indeed leads to Raychaudhuri's equation (since $\rho^{\frac{2}{n}-1}\neq 0$). This generalization is valid in general $(n+1)$-dimensions and reduces to the 
expressions discussed in \cite{alsa} for $n=2$.

Actually, the above calculation depends on a simplification that is not valid in general. Equation (\ref{raycl}) tells us that $V[\rho]$ is actually a functional, and not a simple function of $\rho$, as $\sigma^2$ contains  $\rho^\prime$ and $\mathcal{R}$ may contain terms having up to the second derivative, $\rho^{\prime\prime}$. However, this calculation  can still be implemented for some 
simple cases, where $\rho$ is a function of a single variable, so that all its derivatives are functions  of that variable and in principle can be written as a function of $\rho$. The simplest example is certainly the spatially isotropic and homogeneous cosmology. We shall show this for a spatially flat universe. The method is also likely to work for some spatially homogeneous anisotropic models like Bianchi I. In all these cases, $\rho$ is a function of the cosmic time $t$ only, and thus  derivatives of $\rho$ can be written as a function of 
$\rho$. Thus the functional $V[\rho]$ can be treated as a function, $V=V(\rho)$. For inhomogeneous cosmologies, the method will not work, as $V$ will remain a functional. The total derivative 
with respect to the affine parameter $\lambda$ in equation (\ref{el-eq}) will lead to partial derivatives with respect to spatial coordinates as well. 

Another more crucial issue needs to be pointed out here. The approach of Alsaleh \textit{et al}\cite{alsa} treats $\rho$ as the dynamical variable and Raychaudhuri's equation, derived from an action, determines the dynamics. But it is important to note that Raychaudhuri's equation (\ref{raych}) is an identity in Riemannian geometry, very much along the lines of  
the Bianchi identities ($G^{\mu \nu}_{;\nu} = 0$). So it does not have any dynamical content of its own. We shall also show that the construction of the potential $V$ clearly indicates this. For a detailed description on similar issues with the Noether currents, derived from a given geometry, we refer to the work of Padmanabhan\cite{paddygrg}.

We shall now see an illustration of the issues mentioned by making an attempt to construct the potential with the simplest example of a spatially flat Friedmann-Robertson-Walker (FRW) metric given by,

\begin{equation}\label{frwf}
 \mathrm{d}s^2=-\mathrm{d}t^2+a^2(t)\left[{\mathrm{d}r^2}+r^2 \mathrm{d}\vartheta^2+r^2 \sin^2\vartheta \mathrm{d}\phi^2\right].
\end{equation}

We have, $\rho=\sqrt{h} =a^3$, can choose $\lambda=t$, and thus get $\mathcal{R}=-3\frac{\ddot{a}}{a}$, where a dot indicates a derivative with respect to $t$, the cosmic
time. We also have $\sigma^2=0$ as the given spacetime is spatially isotropic. Therefore,
\begin{equation}
 \rho^\prime=\frac{\mathrm{d}\rho}{\mathrm{d}t}\equiv \dot{\rho}=3a^2 \dot{a}.
\end{equation}

From equation \eqref{delta-v1} we have,
\begin{equation}\label{potc}
 \delta V=-9\ddot{a}\delta a=\delta\left(\frac{9}{2}\dot{a}^2\right)-\frac{\mathrm{d}}{\mathrm{d}t}\left(9\dot{a}\delta{a}\right).
\end{equation}

If we invoke the condition, used in a standard variational principle, that there is no variation on the boundary, we can ignore the total derivative term, and identify $V$ as 
 
\begin{equation}
V=\frac{9}{2}\dot{a}^2,
\end{equation}
and the Lagrangian \eqref{lagr} is,
\begin{equation}
 L=\frac{9}{2}\dot{a}^2-\frac{9}{2}\dot{a}^2=0.
\end{equation}

This triviality is a consequence of the fact that the Raychaudhuri's equation is an identity in Riemannian geometry as already pointed out. If $\mathcal{R}$ is expressed in terms of the metric components ($a$) and its derivatives, one arrives at a triviality.

A possible way out for a somewhat meaningful interpretation of this Lagrangian formulation is the following. 
In General Relativity we can use the field equations to represent $\mathcal{R}=R_{\mu\nu}u^\mu u^\nu$ in terms of the energy-momentum tensor, $T_{\mu\nu}$. Therefore, we write from equation \eqref{delta-v1},
\begin{equation}\label{delta-v2}
 \frac{\partial V}{\partial \rho}=\rho^{\left(\frac{2}{n}-1\right)}\left(2\sigma^2+\kappa T_{\mu\nu}u^\mu u^\nu+\frac{\kappa}{2}T\right),
\end{equation}
where $T$ is the trace of the energy-momentum tensor. We have used Einstein's equations,
\begin{equation}
 R_{\mu\nu}-\frac{1}{2}g_{\mu\nu}R=\kappa T_{\mu\nu},
\end{equation}
and continue with the assumption that all variables are functionally related to $\rho$ so that $V$ can be treated as a function of $\rho$. Here $R$ is the Ricci scalar and $\kappa=8\pi G$ with $G$ being the Newtonian gravitational constant.
We can bypass the triviality by proceeding in this way and go on to construct a non-trivial Lagrangian. We will illustrate this point with the example already considered - namely spatially flat FRW universe.

We will consider a scale factor with power law dependence on $t$. For a universe described by an FRW metric containing a distribution of perfect fluid having an equation of state 
$p=w \epsilon$, where $p, \epsilon$ are the pressure and density of the fluid with $w$ being a constant, one has solution for $a$ as,
\begin{equation}\label{scalef}
a=a_0 (t-t_0)^{C},
\end{equation}
where $C, a_0, t_0$ are constants. Here, it follows from equation \eqref{delta-v2} that,
\begin{equation}\label{delta-v3}
 \frac{\partial V}{\partial \rho}=\rho^{-\frac{1}{3}}\frac{\kappa}{2}(\epsilon+3p).
 \end{equation}
 
 For the scale factor as given by equation \eqref{scalef}, we have,
 \begin{equation}
 \frac{\kappa}{2}(\epsilon+3p)=\frac{\tilde {D}}{(t-t_0)^2}= \frac{D}{a(t)^{\frac{2}{C}}}=\frac{D}{\rho^{\frac{2}{3C}}},  
 \end{equation}
 where $\tilde{D}$, and hence $D$, are constants.
Now, the potential can be obtained from equation \eqref{delta-v3} as,
\begin{equation}
 V=k \rho^{\frac{2}{3}(1-\frac{1}{C})},
\end{equation}
where $k$ is a constant. Thus, the Lagrangian is given by (using equation \eqref{lagr}),
\begin{equation}
 L=\frac{1}{2}\rho^{-\frac{4}{3}}\dot{\rho}^2-k \rho^{\frac{2}{3}(1-\frac{1}{C})}.
\end{equation}

We have used the functional dependence of $\rho$ (through $a$) on $t$ to express $V$ as a function of $\rho$.

Another example where we can construct the potential $V$ and thus the Lagrangian following the similar procedure is,
\begin{equation}
 a(t)=\left[A \exp(\lambda t)+B \exp (-\lambda t)\right]^{\frac{2}{3}}.
\end{equation}
This is the general solution for the scale factor when we assume that the jerk parameter of the universe,
$ j\equiv\left(\frac{a}{\dot{a}}\right)^3\frac{\dddot{a}}{a}= 1$\cite{sgc}. 

{Even more complicated cases, where the scale factor or the variables in the matter sector cannot be written as a function of the cosmic time $t$ explicitly, 
can also be dealt with, such as the one recently given by Shokri {\it et al}\cite{shokri}, where one or more of the quantities like the scale factor $a$ (hence $\rho$ in the present context), 
the scalar field $\phi$ and the potential $U=U(\phi)$ are known as a function of $t$ only implicitly. This is possible as in the spatially homogeneous models, there is essentially 
only one independent variable, namely the cosmic time $t$.} 

This procedure can also be extended to spatially anisotropic but homogeneous models where $\rho$ can be expressed as a function of $t$ alone. For an inhomogeneous cosmological model, this simple method will not work, as all the quantities cannot be expressed as a function of $\rho$.

We now proceed to construct the Hamiltonian, acknowledging that the method has only a very limited domain of application. 

The canonically conjugate momentum corresponding to $\rho$ is,
\begin{equation}
 \Pi=\frac{\partial L}{\partial {\rho ^\prime}}=\rho^{\left(\frac{2}{n}-2\right)}{\rho ^\prime}=\rho^{\left(\frac{2}{n}-1\right)}\theta,
\end{equation}
and the Hamiltonian is given by,
\begin{equation}\label{Hamil}
 H=\frac{1}{2}\rho^{\left(2-\frac{2}{n}\right)}\Pi^2+V[\rho],
\end{equation}

The Hamiltonian's equations of motion are,
\begin{equation}
 {\rho ^\prime}=\theta \rho,
\end{equation}
and
\begin{equation}
  \rho^{\left(\frac{2}{n}-1\right)}\left[{\theta ^\prime}+\frac{1}{n}\theta^2+\mathcal{R}+2\sigma^2\right]=0,
\end{equation}
as expected.

\section{Canonical Quantization}\label{sec4}

For a canonical quantization of the system under consideration, $\rho$ and $\Pi$ are promoted to operators such that they satisfy the canonical commutation relation,
\begin{equation}
 [\hat{\rho},\hat{\Pi}]=i\hbar.
\end{equation}

These operators act on the geometric flow state $\Psi[\rho,\lambda]$\cite{alsa}. In $\rho$-representation, we have,
\begin{equation}
 \hat{\rho}=\rho, \hspace{0.2cm} \hat{\Pi}=-i\hbar\frac{\partial}{\partial \rho}.
 \end{equation}
  
The Hamiltonian \eqref{Hamil}, in terms of the operators, is given by 
 \begin{equation}\label{qham1}
\hat{H}=  -\frac{\hbar^2}{2}\rho^{\left(2-\frac{2}{n}\right)}\frac{\partial^2}{\partial\rho^2}+V[\rho].
 \end{equation}

Therefore, the evolution equation for the geometric flow state $\Psi$ can be written as,
\begin{equation}\label{sleq}
 \hat{H}\Psi=i\hbar \frac{\partial}{\partial\lambda}\Psi.
\end{equation}
This is the equation which dictates the evolution of a timelike geodesic congruence in the quantized version and is thus the quantum analogue of the classical Raychaudhuri 
equation, which is applicable to only a limited class of geometries, but definitely for all $n$. 

One expected field of application of this quantized Raychaudhuri equation is the quantized quantum cosmological models. 
Such models, particularly the spatially anisotropic 
ones in the Wheeler-DeWitt quantization scheme\cite{wheeler, dewitt} can pose the problem of a non-unitary evolution\cite{pinto, wilt, hall, alvarenga}.  It has been shown that 
with a proper choice of operator ordering the models can have unitary evolution\cite{barun, sridip1, sridip2, sridip3}. It was also shown that, at least for spatially homogeneous 
models, a self-adjoint extension is always possible\cite{sridip4}. In the 
same way one can show that as the operator (\ref{qham1}) is indeed a symmetric operator with the norm defined as,
 $\left|\left|\Psi\right|\right|=\int_0^\infty d\rho \rho^{\left(\frac{2}{n}-2\right)} \Psi^* \Psi$, the self-adjoint extension is guaranteed via Friedrichs 
 theorem (see \cite{sridip4} and \cite{reed}). 

This can be shown more explicitly as follows.  We choose the operator ordering as, 
 \begin{equation}
  \hat{H}=-\frac{\hbar^2}{2}\rho^{\left(1-\frac{1}{n}\right)}\frac{\partial}{\partial \rho}\rho^{\left(1-\frac{1}{n}\right)}\frac{\partial}{\partial \rho}+V[\rho].
 \end{equation}
With the change of variable,
\begin{equation}
 \chi=n\rho^{\frac{1}{n}},
\end{equation}
one can write,
\begin{equation}\label{qham2}
 \hat{H}=-\frac{\hbar^2}{2}\frac{\partial^2}{\partial \chi^2}+V[\chi],
\end{equation}
which is manifestly symmetric with the norm,
\begin{equation}
\left|\left|\Psi\right|\right|=\int_0^\infty d\chi \Psi^* \Psi.
\end{equation}
 
Therefore, the Hamiltonian admits a self-adjoint extension. This resolves the apprehension of arriving at an equation that 
dictates a non-unitary evolution of geodesic congruences.

\section{Conclusion}\label{sec5}

The work of Alsaleh {\it et al}\cite{alsa} on representing a geodesic congruence as a dynamical system  and quantizing the evolution of a 
timelike geodesic congruence in an $(n+1)$-dimensional spacetime is critically discussed in the present work. Although this kind of quantization is an important endeavour, there are several serious issues. The method apparently looks very promising, but it turns out that it  is not  generally applicable. The basic reason behind this is that Raychaudhuri's equation is actually an identity in Riemannian geometry, and naturally is not one to be obtained from a variational principle as the equation of motion for a geodesic congruence.

We have shown that where $\rho = \sqrt{h}$ is a function of only one variable, for instance the cosmic time $t$, one can express $\rho$ and its derivatives all as functions of $t$. Thus we 
assume that $\rho$ and its derivatives are ``functionally related''. This  helps in writing the functional $V[\rho]$ as a simple function $V(\rho)$. This simplifies 
the problem so that one can construct the ``correct'' Lagrangian using Helmholtz conditions regarding the inverse problem of the Lagrangian formulation. We have constructed the classical Hamiltonian for the system and quantized the system using canonical commutation relations. This is valid for an arbitrary $n+1$ dimensions, and the results  reduce to those of \cite{alsa} for $n=2$.

This simple method will, as we have shown, lead to a triviality, consistent with the fact that Raychaudhuri's equation is a geometric identity. Now if we try to construct the potential 
using $R_{\mu\nu} u^\mu u^\nu$ through Einstein's field equations, we can arrive at some non-trivial results. By using Einstein's equations, Raychaudhuri's equation is no longer an identity of 
Riemannian geometry, and thus one can bring some meaning out of the Lagrangian formulation.

We then write down the evolution equation for the quantized geodesic congruence in terms of the geometric flow state $\Psi$. In the quantum picture this equation plays the same role that the Raychaudhuri equation does in the classical one. 

It definitely deserves mention that all the limitations that we are talking about, are present at the classical level itself. The quantization scheme is quite straightforward, and does not lead to any additional issue.

The present work has an additional feature. We checked that the self-adjoint extension of the Hamiltonian is quite possible in this context. This ensures the conservation of probability. This is in fact a crucial issue for the quantum description of any physical system. 

One should note that some of the important results of the equation derived by Alsaleh \textit{et al}\cite{alsa} are retained in the ``corrected'' version. In the expression of the 
Hamiltonian \eqref{qham1} the effective mass is given by,
\begin{equation}
 m_{\mbox{eff}}=\rho^{\left(\frac{2}{n}-2\right)}.
\end{equation}
For $n>1$ this mass diverges as $\rho\rightarrow 0$ leading to an infinite effective potential, and a proper boundary condition would be $\Psi(\rho = 0) = 0$ 
(see \cite{avijit} for a comprehensive description). Thus, the probability of focusing of the congruence (implemented by $\rho\rightarrow 0$) is vanishingly small. 

This analysis can also be extended to null geodesics in arbitrary dimensions ($n>1$). For this, one has to replace $n$ by $n-1$ from the beginning so as to be consistent 
with the Raychaudhuri equation for null geodesics\cite{poisson, burg}.

This quantization of the geometric flow should be useful in understanding the existence of singularities in gravitational systems. As already mentioned, 
the focusing of time-like geodesics in spatially homogeneous cosmological systems without vorticity can be studied with the help of the equation developed in this work. 
For instance, all Friedmann models and most of the anisotropic Bianchi models can be examined for the singularities in the quantum regime. 

An earlier significant attempt towards a quantum version of the Raychaudhuri equation was by Das, where the classical geodesics were replaced by Bohmian trajectories\cite{saurya}.

An arena, where the quantized Raychaudhuri equation should find an immediate application is black hole physics. The modification of the Raychaudhuri 
equation in \cite{saurya} found an immediate application in the estimation of the Hawking temperature\cite{vagenas}, {where corrections to Hawking temperature of a 
Schwarzchild black hole are obtained from the Quantum Raychaudhuri equation, and compared with that obtained using a ``linear'' correction and a ``linear plus quadratic'' 
correction arising from the Generalized Uncertainty Principle. It was found that all of them tend to prevent a catastrophic evaporation of the black hole.} 
Very recently there is an attempt towards a resolution of the black hole singularity in loop quantum gravity where a modified Raychaudhuri equation has been used\cite{saurya1}.

In the present work, the classical geodesic flows are not replaced but rather, are quantized themselves. This is expected to find application in the investigation of the 
singularities in the quantum regime for a collapse of homogeneous systems, such as the Datt-Oppenheimer-Snyder collapse\cite{datt, oppen}.

\section*{Acknowledgements}
Shibendu Gupta Choudhury thanks Council of Scientific and Industrial Research, India for the financial support. The authors are grateful to Thanu Padmanabhan for his enlightening comments on an earlier draft, pointing out some major issues, which helped us improving the draft to a large extent. Thanks are also due to Krishnakanta Bhattacharya for helping us (with some suggestions) on the draft. 

%%%%%%%%%%%%%%%%%%%%%%%    BIBLIOGRAPHY   %%%%%%%%%%%%%%%%%%%%%%%%

\bibliographystyle{elsarticle-num}

\begin{thebibliography}{99}
\bibitem{penrose} R. Penrose, Phys. Rev. Lett. {\bf 14},  57 (1965).

\bibitem{hawking} S. W. Hawking and R. Penrose, Proc. Roy. Soc. Lond. A {\bf 314}, 529 (1970).

\bibitem{niz} G. Niz and N. Turok, Phys. Rev. D {\bf 75}, 026001 (2007).

\bibitem{leh1} J. L. Lehners, P. McFadden and N. Turok, Phys. Rev. D {\bf 75}, 103510 (2007).

\bibitem{leh2} J. L. Lehners and N. Turok, Phys. Rev. D {\bf 77}, 023516 (2008).

\bibitem{green} B. Greene, D. Kabat and S. Marnerides, Phys. Rev. D {\bf 80}, 063526 (2009).

\bibitem{cope} E. J. Copeland, G. Niz and N. Turok, Phys. Rev. D {\bf 81}, 126006 (2010).

\bibitem{casa} R. Casadio, Int. J. Mod. Phys. D {\bf 9}, 511 (2000).

\bibitem{lang} P. Laguna, Phys. Rev. D {\bf 75}, 024033 (2007).

\bibitem{batt} M. V. Battisti and A. Marciano, Phys. Rev. D {\bf 82}, 124060 (2010).

 \bibitem{asht0} A. Ashtekar, T. Pawlowski and P. Singh, Phys. Rev. Lett. {\bf 96}, 141301 (2006).

\bibitem{agu} I. Agullo, A. Ashtekar and W. Nelson, Class. Quantum Grav. {\bf 30}, 085014 (2013).

\bibitem{awad} A. Awad and A. F. Ali, J. High Energy Phys. {\bf 1406}, 093 (2014).

\bibitem{trug} C. A. Trugenberger, Phys. Rev. D {\bf 92}, 084014 (2015).

\bibitem{gara} R. Garattini and M. Faizal, Nucl. Phys. B {\bf 905}, 313 (2016).

\bibitem{oriti} D. Oriti, L. Sindoni and E. W. Ewing, Class. Quantum Grav. {\bf 34}, 04LT01 (2017).

\bibitem{hof} S. Hofmann and M. Schneider, Phys. Rev. D {\bf 91}, 125028 (2015).

\bibitem{fermilab} Fermi-LAT Collab. (M. Ajello et al.), J. Cosmol. Astropart. Phys. {\bf 1202}, 012 (2012).

\bibitem{rosz} L. Roszkowski, E. M. Sessolo and A. J. Williams, J. High Energy Phys. {\bf 1408}, 067
(2014).

\bibitem{nicol} H. Nicolai, K. Peeters and M. Zamaklar, Class. Quantum Grav. {\bf 22}, R193 (2005).

\bibitem{rc} A. K. Raychaudhuri, Phys. Rev. {\bf 98}, 1123 (1955).

\bibitem{ehlers} J. Ehlers, Akad. Wiss. Lit. Mainz, Abhandl. Math.-Nat. Kl. {\bf 11}, 793 (1961); translation: J. Ehlers, Gen. Relativ. Gravit. {\bf 25}, 1225 (1993).

\bibitem{saurya} S. Das, Phys. Rev. D {\bf 89}, 084068 (2014).

\bibitem{ali1} A. F. Ali and S. Das, Phys. Lett. B {\bf 741}, 276 (2015).

\bibitem{ali2} A. F. Ali and M. M. Khalil, Nucl. Phys. B {\bf 909}, 173 (2016).

\bibitem{bojo} M. Bojowald, Phys. Rev.
Lett. {\bf 86}, 5227 (2001).

\bibitem{asht} A. Ashtekar, J. Phys. Conf. Ser. {\bf 189}, 012003 (2009). 

\bibitem{singh} P. Singh, Class. Quantum Grav. {\bf 26}, 125005 (2009).

\bibitem{li} L.-F. Li and J.-Y. Zhu, Adv. High Energy Phys. {\bf 2009}, 905705 (2009).

\bibitem{ashtbojo} A. Ashtekar and M. Bojowald, Class. Quantum Grav. {\bf 23}, 391 (2006).

\bibitem{burg} D. J. Burger, N. Moynihan, S. Das, S. S. Haque, and B. Underwood, Phys. Rev. D
{\bf 98}, 024006 (2018).

\bibitem{moti} R. Moti and A. Shojai, Phys. Rev. D {\bf 101}, 064013 (2020). 

\bibitem{sumanta} S. Chakraborty, D. Kothawala and A. Pesci, Phys. Lett. B {\bf 797}, 134877 (2019).

\bibitem{kothaw} D. Kothawala, Phys. Rev. D {\bf 88}, 104029 (2013).

\bibitem{paddykotha} D. Kothawala and T. Padmanabhan, Phys. Rev. D {\bf 90}, 124060 (2014).

\bibitem{alsa} S. Alsaleh, L. Alasfar, M. Faizal and A. F. Ali, Int. J. Mod. Phys. A {\bf 33}, 1850052 (2018).


\bibitem{davis1} D. R. Davis, Trans. Amer. Math. Soc. {\bf 30}, 710 (1928).

\bibitem{davis2} D. R. Davis, Bull. Amer. Math. Soc. {\bf 35}, 371 (1929).

\bibitem{douglas} J. Douglas, Trans. Amer. Math. Soc. {\bf 50}, 71 (1941).

\bibitem{casetta} L. Casetta and C. P. Pesce, Acta Mechanica {\bf 225}, 1607 (2014).

\bibitem{cramp} M. Crampin, T. Mestdag and W. Sarlet, Z. Angew. Math. Mech. {\bf 90}, 502 (2010).

\bibitem{nigam} K. Nigam and K. Banerjee, arXiv:1602.01563.

\bibitem{poisson} E. Poisson, {\it A Relativist's Toolkit: The Mathematics of Black-Hole Mechanics} (Cambridge University Press, Cambridge 2004).

\bibitem{paddygrg} T. Padmanabhan, Gen. Relativ. Gravit. \textbf{46}, 1673 (2014).

\bibitem{sgc} S. G. Choudhury, A. Dasgupta and N. Banerjee,  Mon. Not. Roy. Astron. Soc. {\bf 485}(4) 5693 (2019).

{\bibitem{shokri} M. Shokri, J. Sadeghi, M. R. Setare and S. Capozziello, Int. J. Mod. Phys. D {\bf 30}(09), 2150070 (2021).}

\bibitem{wheeler} J. A. Wheeler, Superspace and the nature of quantum geometrodynamics in {\it Battelle Rencontres} (Benjamin, New York 1968).

\bibitem{dewitt} B. S. DeWitt, Phys. Rev. {\bf 160}, 1113 (1967).

\bibitem{pinto} N. Pinto-Neto and J. C. Fabris, Class. Quantum Grav. {\bf 30}, 143001 (2013).

\bibitem{wilt} D. L. Wiltshire, “Cosmology: The Physics of the Universe”, in {\it Proceedings of the 8th Physics Summer School}, Vol. 16, pp. 473–531 (Australian National University, Canberra, Australia 1996).

\bibitem{hall} J.  J.  Halliwell, in {\it Quantum  Cosmology  and  Baby  Universes},  edited  by  S.  Coleman,  J.  B.  Hartle,  T.  Piran  and  S.Weinberg (World Scientific, Singapore 1991).

\bibitem{alvarenga} F. G. Alvarenga, A. B. Batista, J. C. Fabris and S. V. B. Goncalves, Gen. Relativ. Gravit. {\bf 35}, 1659 (2003).

\bibitem{barun} B. Majumder and N. Banerjee, Gen. Relativ. Gravit. \textbf{45}, 1 (2013).

\bibitem{sridip1} S. Pal and N. Banerjee, Phys. Rev. D {\bf 90}, 104001 (2014).

\bibitem{sridip2} S. Pal and N. Banerjee, Phys. Rev. D {\bf 91}, 044042 (2015).

\bibitem{sridip3} S. Pal and N. Banerjee, Class. Quantum Grav. {\bf 32}, 205005 (2015).

\bibitem{sridip4} S. Pal and N. Banerjee, J. Math. Phys. {\bf 57}, 122502 (2016).

\bibitem{reed} M. Reed and B. Simon, \textit{Methods of Modern Mathematical Physics}, Vol. 2, 2nd ed. (Academic Press Inc.,
New York 1975). 

\bibitem{avijit} A. Chowdhury and N. Banerjee, Phys. Rev. D {\bf 102}, 124051 (2020).

 \bibitem{vagenas} E. C. Vagenas, L. Alasfar, S. M. Alsaleh and A. F. Ali, Nucl. Phys. B {\bf 931}, 72 (2018).

\bibitem{saurya1} K. Blanchette, S. Das, S. Hergott and S. Rastgoo, Phys. Rev. D {\bf 103}, 084038 (2021).

\bibitem{datt} B. Datt, Z. Phys. {\bf 108}, 314 (1938).

\bibitem{oppen} J. R. Oppenheimer and H. Snyder, Phys. Rev. {\bf 56}, 455 (1939).

\end{thebibliography}

\end{document}